\documentclass[showpacs,preprintnumbers,amsmath,amssymb,axodraw]{revtex4}

\usepackage{amsmath}
\usepackage{graphicx}
\usepackage{epsf}
\usepackage{psfrag}
\usepackage{epsfig}
\usepackage{graphics}
\usepackage[dvips]{color}

\begin{document}

\newcommand{\be}{\begin{equation}}
\newcommand{\ee}{\end{equation}}
\newcommand{\bq}{\begin{eqnarray}}
\newcommand{\eq}{\end{eqnarray}}

\title{Dual embedding of extended models with a Lorentz-breaking mass term}

\date{\today}

\author{H. G. Fargnoli$^{(a)}$} \email[]{helvecio@dex.ufla.br}
\author{L. C. T. Brito$^{(a)}$} \email[]{lcbrito@dex.ufla.br}
\author{A. P. Ba\^eta Scarpelli$^{(b)}$} \email[]{scarpelli.apbs@dpf.gov.br}
\author{Marcos Sampaio$^{(c),(d)}$} \email []{marcos.sampaio@durham.ac.uk}

\affiliation{(a) Universidade Federal de Lavras - Departamento de Ci\^encias Exatas \\
Caixa Postal 3037, 37.200-000, Lavras, Minas Gerais, Brazil}

\affiliation{(b)Setor T\'ecnico-Cient\'{\i}fico - Departamento de Pol\'{\i}cia Federal \\
Rua Hugo D'Antola, 95 - Lapa - S\~ao Paulo}

\affiliation{(c) Departamento de F\'{\i}sica - ICEx - Universidade Federal de Minas Gerais\\ P.O. BOX 702, 30.161-970, Belo Horizonte - MG - Brasil}

\affiliation{(d)Centre for Particle Theory, Department of Mathematical Sciences, Durham University, South Road Durham DH1 3LE, UK}

\begin{abstract}

\noindent
In this paper, we investigate a CPT-even model with a Lorentz-violating mass term. Such kind of models may present very interesting features like superluminal modes of propagation or even instantaneous long-range interactions. The mass term we investigate can be induced at classical or quantum level in a scenario with spontaneous gauge symmetry breaking in a gauge-Higgs model without Lorentz symmetry. We analyze the physical consistency of the model by studying the poles of the propagator. By using the Noether Dualization Method (NDM), we obtain a gauge invariant dual theory for this model. The physical equivalence between the two models is investigated and a general relation between the two propagators is obtained.

\end{abstract}


\maketitle

\section{Introduction}
The Standard Model Extension (SME) \cite{kostelecky1}-\cite{coleman2} provides a description of Lorentz and CPT violation in Quantum Field Theories controlled by a set of coefficients whose small magnitudes are in principle fixed by experiments. The experimental results set stringent bounds in such coefficients, as can be found in the collection of data contained in \cite{data-exp}. Nevertheless, many efforts have been dedicated to these models in the search of more fundamental theories beyond the Standard Model.

The first model which called the attention of the researchers was proposed by Carroll, Field and Jackiw \cite{Carroll}. This model includes a Chern-Simons-like term which violates Lorentz and CPT symmetries due to the presence of a constant background vector that selects a preferred direction in spacetime. However, astrophysical data establish stringent bounds to this kind of vacuum birefringence \cite{Carroll},\cite{Goldhaber}. A question which arose is whether or not this term could be radiatively induced if a Lorentz- and CPT-violating axial term is included in the fermionic sector. The result, which is regularization dependent, has been obtained by many different approaches (see, for example, \cite{CS1}-\cite{CS8}).

Besides the Carroll-Field-Jackiw (CFJ) term, concerning the gauge sector, the SME encompasses a CPT-even one, which is controlled by a constant fourth-rank background tensor with the same symmetries of the Riemann tensor.  The radiative generation of a particular form of the CPT-even term \cite{Carroll2} has been studied in effective models which include Lorentz-violating nonminimal couplings \cite{NM-coupling}.
These nonminimal couplings have been classically studied in several papers \cite{belich-mmferreira}, while more general dimension-five operators have been considered in \cite{shan-quan1}, \cite{kost-5d} and \cite{shan-quan2}.

An interesting aspect in relativity-breaking models to which it has not been given much attention is the possible presence of Lorentz-violating mass terms. Some very interesting aspects were pointed out in \cite{lv-mass1} and \cite{lv-mass2}, where a mass term of the type $-(1/2)m^2 A_j A^j$ in electrodynamics was considered, being $j$ a spatial index. In this case, the gauge field has two massive degrees of freedom, but the static force between charged particles is Coulomb-like. In other words, we have massive propagating degrees of freedom, but we also have instantaneous long-range interactions. In \cite{lv-mass3}, it was considered the extended QED with a Lorentz- and CPT-violating axial term in the fermion sector. It was pointed out that at second order in the background vector $b_\mu$ it is possible to radiatively generate a Lorent-breaking mass for the photon. It was carried out an analysis of more general mass terms, showing the possibility of existence of superluminal modes in such cases. In \cite{lv-mass4} a Stueckelberg lagrangian for massive photons in a generalized $R_\xi$ gauge was studied, with focus in perturbative features of quantum calculations.

Lorentz-breaking mass terms can also be generated by spontaneous gauge symmetry breaking in a Lorentz-violating gauge-Higgs model \cite{lv-mass5}, coming from the symmetric part of the second-rank background tensor which couples to the kinetic part of the Higgs field. It can also emerge, along with a CPT-even aether term for the gauge sector, by quantum corrections in a gauge-Higgs model with a Carroll-Field-Jackiw term \cite{LCTBrito}. We are particularly interested in the kind of mass term which has been quantically induced in \cite{LCTBrito}. Since the aether term of \cite{Carroll2} is induced along with this mass term, it will also be considered here. Classically, the form of the relativity-breaking mass term we are interested in can also emerge from the Higgs mechanism of \cite{lv-mass5} if we consider a particular form of the second-hank tensor of the Higgs sector.

Concerning the class of models described above, we are also interested in investigating how the gauge embedding works with such kind of gauge symmetry violation. Since one can consider the non-invariant model as the gauge fixed version of a gauge theory, it would be useful to carry out such analysis. Hidden symmetries may be revealed by the construction of a gauge invariant theory from a non-invariant one. In other words, one model would reduce to the other under some gauge fixing conditions.

The concept of duality between two different models in field theory  is very interesting and useful, since it allows for the mutual mapping of theories possessing essentially different actions. There are some important features that are manifest in one model but are hidden in the other one. The duality was first established in three spacetime dimensions in the case of self-dual and Maxwell-Chern-Simons models \cite{Deser-Jackiw}, \cite{Kenneth}.

One approach to determine the physical equivalence between two theories is the master action procedure \cite{Deser-Jackiw}, \cite{review}, which is constructed starting with the self-dual model and then adding a mixing term in  the two fields. The two models can be obtained from the master action by using the equation of motion of one of the fields in the original action.

On the other hand, the gauging iterative Noether Dualization Method (NDM) \cite{Ilha1} has been shown to be effective in establishing dualities between some models \cite{Ilha2}. This approach is based on the idea of a local lifting of a global symmetry and is accomplished by an iterative embedding of Noether counterterms. The idea of the procedure may be traced back to the papers by Freedman and van Nieuwenhuizen \cite{Nieuw} and subsequent works by Ferrara,
Freedman and van Nieuwenhuizen \cite{Ferrara1} and Ferrara and Scherk \cite{Ferrara2}, which were important for the construction of component-field supergravity actions. In our context, this method provides a strong suggestion of duality, since it furnishes the expected result in the paradigmatic duality between the self-dual and Maxwell-Chern-Simons models in three dimensions.

The Noether Dualization Method has been applied in the context of Lorentz-violating models with interesting results \cite{Bota}, \cite{Petrov}. In \cite{Bota}, the Carrol-Field-Jackiw model with a Procca term was dualized. The intriguing result is that, although the two dual models share the same physical spectrum, the gauge theory resulting from the NDM procedure acquires ghost modes, which are indeed originated from the original CFJ model without the Procca mass term. This interesting fact has been shown to be a general result when NDM is applied to Procca-like models \cite{scarp-epl}. This result is made explicit by the general relation between the propagators of the dual models.

Alternatives to avoid the emergency of ghosts in the process of dualization were developed \cite{Dalmazi1}, \cite{Dalmazi2}, \cite{Dalmazi3}, \cite{Dalmazi4}. In some approaches, the price to be paid is the lost of locality \cite{Dalmazi1}. For the case of a spin-2 self-dual field in three spacetime dimensions, it was shown that the dual model achieved by NDM does not suffer with the presence of ghosts \cite{Dalmazi4}.

In this paper, we carry out an analysis of a CPT-even Lorentz-violating model with a mass term of the type $-(1/2)m^2(g_{\mu \nu} - \beta b_\mu b_\nu)$ and with the kinetic aether-like term of \cite{Carroll2}, $- \frac \rho2 \left(b_\mu F^{\mu \nu}\right)^2$. We use the dimensionless parameters $\beta$ and $\rho$ to discuss, for example, in what kind of situation we could have superluminal modes of propagation. Particular attention is paid to the physical spectra of the theories and to how the degrees of freedom are distributed amongst the physical modes. We also carry out an analysis of the process of dualization by means of Noether embedding in the context of models with Lorentz-violating mass terms. We are mainly interested in issues such how the relation between the propagators of the models is modified and to what extension these models can be considered equivalent.

In section II, we present the class of models which we are interested in and analyze their spectra for some different combinations of the values of the coefficients of the Lorentz-breaking terms. The NDM is used in order to derive the dual model for this class of theories in section III, where we also compare its spectrum with the original one. In section IV, it is carried out a general discussion on the Noether embedding of models with Lorentz-violating mass. We conclude in section V.

\section{A CPT-even model with a Lorentz-breaking mass}

As the origin of our model at classical level, we present the gauge-Higgs theory,
\bq
\label{higgsmodel}
&&{\cal L}_{gH}=-\frac 14 F_{\mu \nu}F^{\mu \nu} - \frac 12 \left(c_\mu F^{\mu \nu}\right)^2 +
\left(g^{\mu \nu}+ k^{\mu \nu}_\phi\right)\left(D_\mu \phi\right)^*\left(D_\nu \phi\right) \nonumber \\
&&+ \mu^2\phi^*\phi -\frac {\lambda}{2} \left(\phi^*\phi\right)^2 ,
\eq
in which $\lambda$ and $\mu$ are positive constants, the covariant derivative is given by $D_{\mu}\phi=\partial_{\mu}\phi+ieA_{\mu}\phi$ and $F_{\mu \nu}=\partial_{\mu}A_{\nu}-\partial_{\nu}A_{\mu}$ is the electromagnetic field strength tensor. The lagrangian density above is invariant under local $U(1)$ gauge transformations and the complex scalar field $\phi$ develops a vacuum expectation value $\langle\phi\rangle_{0}=\frac{\mu}{\sqrt{\lambda}}\equiv v$, since the $U(1)$ symmetry is spontaneously broken. Rewriting the lagrangian (\ref{higgsmodel}) in terms of real scalar fields $h$ and $\varphi$, such that $\phi=2^{-\frac{1}{2}}\left(h+v+i\varphi\right)$, yields for the pure gauge sector:
\be
{\cal L}=-\frac 14 F_{\mu \nu}F^{\mu \nu} - \frac 12 \left(c_\mu F^{\mu \nu}\right)^2
+ \frac {1}{2}\left(g^{\mu \nu}+ k^{\mu \nu}_S\right)m^2 A_\mu A_\nu,
\ee
with $k^{\mu \nu}_S$ being the symmetric part of $k^{\mu \nu}_\phi$ and $m^2=(\sqrt{2}e v)^2$. If we particularly set $k^{\mu \nu}_S=-\beta b^\mu b^\nu$, with $b^\mu$ being a constant background vector, we match the form of Lorentz breaking mass radiatively induced in \cite{LCTBrito}. Besides, let us consider that we have only one preferred spacetime direction, such that $c^\mu \propto b^\mu$. Under these conditions, we have the following class of CPT-even lagrangian densities for the photon sector
\be
{\cal L}=-\frac 14 F_{\mu \nu}F^{\mu \nu} - \frac \rho2 \left(b_\mu F^{\mu \nu}\right)^2 + \frac {m^2}{2}A^\mu A^\nu h_{\mu \nu},
\label{Proccamunu}
\ee
with
\be
h_{\mu \nu}=g_{\mu \nu}-\beta b_\mu b_\nu,
\ee
where $\rho$ and $\beta$ are dimensionless parameters.

We proceed now to the analysis of this model. After some partial integrations (the lagrangian density is supposed to be under integration), we can write
\bq
&&{\cal L}=\frac 12 A^\mu \left\{ \left( \Box +m^2+\rho\lambda^2 \right)\theta_{\mu \nu} + \left(m^2 +\rho \lambda^2\right) \omega_{\mu \nu} \right. \nonumber \\
&&\left. +\left(\rho \Box -\beta m^2 \right) \Lambda_{\mu \nu} - \rho\lambda\left( \Sigma_{\mu \nu} +\Sigma_{\nu \mu} \right)\right\}A^\nu,
\eq
with $\theta_{\mu \nu}=g_{\mu \nu}-\frac {\partial _\mu \partial _\nu}{\Box}$
and $\omega_{\mu \nu}=\frac {\partial _\mu \partial _\nu}{\Box}$ being the transversal
and the longitudinal operators, respectively, and
\bq
\Lambda _{\mu \nu}&=& b_\mu b_\nu \,\,\,\,\,\,\mbox{and}  \\
\Sigma _{\mu \nu}&=& b_\mu \partial _\nu,
\eq
generated by the inclusion of the external vector $b^\mu$ ($\lambda$ stands for
$\Sigma _\mu \,\,^\mu=b_\mu \partial ^\mu$). The Lorentz algebra of these operators
is shown in Table 1:
\begin{center}
\begin{tabular}{|c|c|c|c|c|c|}
\hline
& $\theta _{\,\,\,\,\,\nu }^{\alpha }$ & $\omega _{\,\,\,\,\,\nu }^{\alpha }$
& $\Lambda _{\,\,\,\,\,\nu }^{\alpha }$ & $%
\Sigma _{\,\,\,\,\,\nu }^{\alpha }$ & $\Sigma_\nu ^{\,\,\,\,\,\alpha}$ \\
\hline
$\theta _{\mu \alpha }$ & $\theta _{\mu \nu }$ & $0$ & $%
\Lambda _{\mu \nu }-\frac{\lambda }{\Box }\Sigma _{\nu \mu }$ & $\Sigma
_{\mu \nu }-\lambda \omega _{\mu \nu }$ & $0$ \\ \hline
$\omega _{\mu \alpha }$ & $0$ & $\omega_{\mu\nu}$ & $\frac{\lambda
}{\Box }\Sigma _{\nu \mu }$ & $\lambda \omega _{\mu \nu }$ & $\Sigma_{\nu
\mu}$ \\ \hline
$\Lambda _{\mu \alpha }$ & $\Lambda _{\mu \nu }-\frac{\lambda }{\Box}%
\Sigma_{\mu \nu }$ & $\frac{\lambda }{\Box}\Sigma_{\mu \nu }$ & $%
b^{2}\Lambda _{\mu \nu }$ & $b^{2}\Sigma _{\mu \nu }$ & $\lambda
\Lambda_{\mu \nu}$ \\ \hline
$\Sigma _{\mu \alpha }$ & $0$ & $\Sigma_{\mu \nu}$ & $\lambda
\Lambda _{\mu \nu }$ & $\lambda \Sigma _{\mu \nu }$ & $\Box \Lambda_{\mu \nu}$ \\ \hline
$\Sigma_{\alpha \mu}$ & $\Sigma_{\nu \mu} -\lambda \omega_{\mu \nu}$ & $\lambda \omega_{\mu \nu}$ & $b^2 \Sigma_{\nu \mu}$ & $b^2 \Box
\omega_{\mu \nu}$ & $\lambda \Sigma_{\nu \mu}$ \\ \hline
\end{tabular}
\vspace{2mm}

Table 1: Multiplicative table fulfilled by $\theta$, $\omega$, $\Lambda$
and $\Sigma$. The products are supposed to obey the order ``row times
column''.
\end{center}

Using the algebra of Table 1 and considering that the Lagrangian density is written in the form
\be
{\cal L}=\frac 12 A^\mu {\cal O}_{\mu \nu} A^\nu,
\ee
we can calculate the propagator, which is
\be
\langle A_\mu A_\nu \rangle =i \left({\cal O}^{-1}\right)_{\mu \nu}.
\ee
We obtain
\bq
&&\langle A_\mu A_\nu \rangle=\frac{i}{A_1 E}\left\{E \theta_{\mu \nu}
+ \frac{1}{m^2}\left[A_1 E +\beta \lambda^2(1+\rho b^2)A_1-(\rho+\beta)\lambda^2 m^2 \right] \omega_{\mu \nu} \right. + \nonumber \\
&&\left. +\left[\beta A_1 -(\rho + \beta)\Box \right] \Lambda_{\mu \nu} + \lambda(\rho + \beta)\left( \Sigma_{\mu \nu} + \Sigma_{\nu \mu}\right)\right\},
\label{completo}
\eq
in which
\be
A_1=\Box+m^2+\rho \lambda^2, \label{A1}
\ee
and
\be
E=\left(1+\rho b^2\right)\Box + m^2\left(1-\beta b^2\right)- \beta \lambda^2(1+\rho b^2). \label{E}
\ee
We are now in position to analyze the physical spectrum of the model for different combinations of values for the constants $\rho$ and $\beta$. We are interested in three situations: the complete model, with $\rho=1$ and $\beta=-1$; the situation where $\rho=1$ and $\beta=0$; and the case in which $\rho=0$ and $\beta=1$. Let us first consider the complete model. We are interested in two situations for the background vector $b_\mu$, namely the cases in which it is space-like or time-like.

\subsection{$b_\mu$ space-like}

We use a representative background vector given by $b^\mu=(0,0,0,t)$. In this case, we have
\be
b \cdot k=-tk^3=t k_3  \,\,\,\,\,\, \mbox{and} \,\,\,\,\,\, b^2=-t^2.
\label{spacelike}
\ee
For the propagator $\langle A_\mu A_\nu \rangle$, we will have in the denominator, in momentum space,
\be
D=A_1 E=(1-\rho t^2)\left(k^2-m^2+ \rho t^2 k_3^2\right) \left[k^2- \frac{(1+\beta t^2)}{(1-\rho t^2)}m^2 -\beta t^2k_3^2\right],
\ee
which gives us poles at
\be
k_{0}^{2} = \bold k^2+ m^2-\rho t^2k_{3}^{2}\equiv m_1^2
\ee
and
\be
k_{0}^{2} = \bold k^2+\beta t^2 k_3^2+\frac{(1+\beta t^2)}{(1-\rho t^2)}m^2\equiv m_2^2.
\ee

It is simple to show, using the momentum space equation of motion for our model,
\be
-\left[k^2+\rho(b \cdot k)^2\right]A_\nu+\left[k_\nu +\rho (b \cdot k)b_\nu\right]k^\mu A_\mu
+\rho\left[(b \cdot k)k_\nu -k^2 b_\nu\right] b^\alpha A_\alpha + m^2 h^\mu\,_\nu A_\mu=0,
\ee
that the pole $m_1^2$ is to be associated to a propagating wave with polarization orthogonal to $b_\mu$, whereas the pole $m_2^2$ is associated to a mode such that
\be
\frac{A_\perp}{A_3}=\frac{(1-\rho t^2)k_3 k_\perp}{m^2+(1-\rho t^2)k_3^2},
\ee
in which $A_\perp$ and $k_\perp$ are the projections of $\bold A$ and $\bold k$ perpendicular to $b_\mu$, respectively. An interesting feature to be observed is the possibility of existence of superluminal modes of propagation. To exemplify, let us take the case where $\rho=0$. In this case, is simple to see that the pole $m_2^2$ corresponds to a superluminal mode if $\beta > 0$.

Our present task consists in checking the features of the poles for $b_\mu$ space-like. Knowing that two different poles appear, we have to study the residue matrix of the vector propagator at each of its poles $k_0^2 = m_1^2$ and $k_0^2=m_2^2$. In order to infer about the physical nature of the simple poles, we have to calculate the eigenvalues of the residue matrix for each of these poles. This is done in the sequel. 

Let us argue that the momentum propagator, $k^\mu$, is actually a Fourier-integration variable and so we are allowed to pick a representative momentum whenever $k^2>0$. We pursue our analysis of the residues by taking $k^\mu=(k^0,0,0,k^3)$. In this analysis, we are interested in checking how the degrees of freedom are distributed amongst the two modes and if they respect physical requests such as unitarity and causality. These features are not expected to vary with the spatial direction of propagation of the electromagnetic wave. Besides, the physics described by the Standard Model Extension (SME) is Lorentz invariant from the observer point of view. For a general passive Lorentz transformation, since the vectors $k^\mu$ and $b^\mu$ are not proportional to each other, they will not remain parallel in the pure spatial sector. So, our choice of $k^\mu$, although very particular, will provide us the answers we are looking for. For this very particular situation, the pole $m_1^2$ corresponds to a transversal mode, whereas the pole $m_2^2$ corresponds to a longitudinal mode. The propagator, for the choices we have made, can be written as
\bq
&&\langle A_\mu A_\nu \rangle= \frac{i}{D} \left\{-(1-\rho t^2)\left(k_0^2-m_2^2\right) \theta_{\mu \nu} +\right. \nonumber \\
&&\left. +\frac{1}{m^2} \left[(1-\rho t^2)\left(k_0^2-m_1^2-\rho t^2 k_3^2\right)\left(k_0^2-m_2^2\right)+ t^2k^2k_3^2\left(\rho +\beta\right)\left(1-\rho t^2\right)\right]\omega_{\mu \nu} +\right.\nonumber \\
&&\left.+t^2\left(\rho k^2+\beta m^2-\beta \rho t^2 k_3^2\right) \delta_\mu^3 \delta_\nu^3 +t^2k_3\left(\rho
+ \beta\right)\left(\delta_\mu^3 k_\nu+ \delta_\nu^3 k_\mu \right) \right\}.
\eq
For the pole $m_1^2$, we find the following residue matrix:
\be
R_{1}=
\left(
\begin{array}{cccc}
0& 0 & 0 & 0 \\
0 & 1 & 0 & 0 \\
0 & 0  & 1  & 0 \\
0 & 0 & 0 & 0
\end{array}
\right),
\ee
with eigenvalues

\bq
\lambda _1 &=& 0,  \\
\lambda _2 &=& 1,  \\
\lambda _3 &=& 1,  \\
\lambda _4 &=& 0.
\eq

As it can be seen, this pole is to be associated with two physical degrees of freedom, since we
have two non-null positive eigenvalues.

We now study the pole $m_2^2$. The associated residue matrix reads:

\be
R_{2}=
\left(
\begin{array}{cccc}
(1+\beta t^2)\frac{k_3^2}{m^2}& 0 & 0 & \frac{|m_2|k_3}{m^2} \\
0 & 0 & 0 & 0 \\
0 & 0  & 0  & 0 \\
\frac{|m_2|k_3}{m^2} & 0 & 0 & \frac{m^2+(1-\rho t^2)k_3^2}{(1-\rho t^2)m^2}
\end{array}
\right),
\ee
with eigenvalues

\bq
\lambda _1 &=& 0,  \\
\lambda _2 &=& 0,  \\
\lambda _3 &=& 0,  \\
\lambda _4 &=& \frac{(2+\beta t^2)(1-\rho t^2)k_3^2+m^2}{(1-\rho t^2)m^2}.
\eq
We have only one degree of freedom associated with this pole. We analyze below the positivity of this eigenvalue for the cases we are interested in.
\begin{itemize}
\item $\rho=1$, $\beta=-1$: this is a very particular and interesting case, since, in this situation, we have $E=(1-t^2)A_1$. It appears we have a dangerous double pole, what could plague the quantum spectrum with ghosts. For this reason, a careful study of this question is worthwhile. However, there occurs a cancelation of one factor of $A_1$ in the denominator in all the sectors of the propagator. We stay with
    \be
    \langle A_\mu A_\nu \rangle = \frac{i}{A_1}\left\{ -\theta_{\mu \nu} + \frac{1}{m^2}\left(k^2-m^2\right) \omega_{\mu \nu}
    + \frac{t^2}{1-t^2} \delta_\mu^3 \delta_\nu^3\right\}.
    \ee
So, it turns out to be a simple pole with three degrees of freedom, although we have a preferred direction in spacetime, since the residue matrix of the propagator has the following eigenvalues:
\bq
&&\lambda_1=0, \nonumber \\
&& \lambda_2=1, \nonumber \\
&& \lambda_3=1, \nonumber \\
&& \lambda_4= \frac{1}{(1-t^2)m^2} \left[m^2+(1-t^2)k_3^2\right].
\eq
A tiny Lorentz violation is characterized by $t^2<<1$, which will assure positive eigenvalues and, therefore, a physical excitation.

\item $\rho=1$, $\beta=0$: in this case, the only Lorentz-violating part is the CPT-even aether term. This tiny Lorentz symmetry breaking causes a little deviation from the spectrum of Procca electrodynamics. We have two massive excitations given by the poles
\be
m_1^2=m^2+(1-t^2)k_3^2
\ee
and
\be
m_2^2=\frac{1}{1-t^2}m^2+k_3^2.
\ee
The analysis of the residue matrices shows that the first pole has two physical degrees of freedom while the second has one, which is physical for $t^2<<1$. The spectrum of a similar model to this one with $\rho=1$ and $\beta=0$ has been analyzed in \cite{Belich-spectrum}.

\item $\rho=0$, $\beta=1$: the deviation of Lorentz symmetry is now realized only by the $b_\mu$ dependent mass term. Again it occurs a tiny modification from the spectrum of the traditional Procca electrodynamics. The degeneracy lifting gives us two massive poles
\be
m_1^2=m^2+k_3^2
\ee
and
\be
m_2^2=(1+t^2)\left(m^2+k_3^2\right),
\ee
being the second one a superluminal mode of propagation.
\end{itemize}

We see, in all the particular situations we studied for the external vector $b_\mu$ space-like, that at tree level the model predicts modes which complies with unitarity (positive norm particles) and causality (positive poles) for $t^2<<1$.

\subsection{$b_\mu$ time-like}

If we instead adopt a time-like Lorentz-violating parameter $b^{\mu}$, with $b^{\mu} = (t,0,0,0)$ and $k^{\mu} = (k^0,0,0,k^3)$, we have
\begin{equation}
b\cdot k = tk_0\qquad \mbox{and}\qquad b^2=t^2.
\end{equation}

Note that, in this case, our choice for the momentum $k^\mu$ allows a general analysis, since the space is isotropic. The poles of our model in this case are
\be
k_{0}^{2} = \frac{k_3^2 + m^2}{(1+\rho t^2)}\equiv \tilde m_1^2
\ee
and
\be
k_{0}^{2} = \frac{k_3^2}{(1-\beta t^2)}+\frac{m^2}{(1+\rho t^2)}\equiv \tilde m_2^2.
\ee
While for the pole $\tilde m_1^2$ the residue matrix is
\be
R_{1}=\frac{1}{1+\rho t^2}
\left(
\begin{array}{cccc}
0& 0 & 0 & 0 \\
0 & 1 & 0 & 0 \\
0 & 0  & 1  & 0 \\
0 & 0 & 0 & 0
\end{array}
\right),
\ee
for the pole $\tilde m^2_2$, we have
\be
R_{2}=
\left(
\begin{array}{cccc}
\frac{k_3^2}{m^2(1-\beta t^2)^2}& 0 & 0 & \frac{|\tilde m_2|k_3}{m^2(1-\beta t^2)} \\
0 & 0 & 0 & 0 \\
0 & 0  & 0  & 0 \\
\frac{|\tilde m_2|k_3}{m^2(1-\beta t^2)} & 0 & 0 & \frac{\tilde m_2^2}{m^2}
\end{array}
\right).
\ee
All the analysis performed for the space-like case can be repeated here, with similar results, the condition for a healthy spectrum being $t^2<<1$.
It is noteworthy the interesting particular case, where $\rho=1$ and $\beta=-1$, in which the spectrum preserves the original degeneracy of the traditional Procca electrodynamics, with only one massive pole with three degrees of freedom.

\section{Noether embedding of the model}

An interesting question which emerges is whether or not it is possible to obtain, from the model discussed in the last section, a gauge invariant physical equivalent theory. We proceed to the gauge embedding of our model. The Noether dualization method consists in a two-step Noether embedding of the gauge symmetry $\delta A_\mu=\partial_{\mu}\eta$ of the two first terms of ${\cal L}$. For this, it is used an auxiliary field $B_\mu$, such that $\delta B_\mu=\delta A_\mu=\partial_{\mu}\eta$, in order to restore gauge symmetry.

Let us then calculate the first variation of our lagrangian density,
\be
\delta {\cal L}[A_{\mu}]\,=
\left\{ \left( \partial^\mu F_{\mu \nu} \right) + \rho b_\mu b^\alpha \left(\partial^\mu F_{\alpha \nu} \right)
-\rho b_\nu b^\alpha \left(\partial^\mu F_{\alpha \mu} \right) + m^2 h_{\mu \nu}A^\mu \right\} \delta A^\nu.
\ee
We may recognize the Noether current as
\be
\label{e2}
J_\nu=\left(\partial^\mu F_{\mu \nu} \right) + \rho b_\mu b^\alpha \left(\partial^\mu F_{\alpha \nu} \right)
-\rho b_\nu b^\alpha \left(\partial^\mu F_{\alpha \mu} \right) + m^2 h_{\mu \nu}A^\mu,
\ee
so that we construct the first iterated lagrangian by introducing an auxiliary field $B$,
\be
{\cal L}^{(1)}\,=\,{\cal L}\,-\,JB.
\ee
Since $B$ transforms as $\delta B_{\mu}\,=\,\delta A_{\mu} = \partial_{\mu} \eta$, then
\be
\delta\,{\cal L}^{(1)}\,=\,-(\delta\,J_{\mu})\,B^{\mu}.
\ee
Using
\be
\delta J^\mu=m^2 h^{\mu \nu} \delta A_\nu,
\ee
we have
\be
\label{varl1}
\delta {\cal L}^{(1)}=-m^2 B_\mu h^{\mu \nu} \delta A_\nu.
\ee
If we define the second iterated lagrangian by
\be
{\cal L}^{(2)}\,=\,{\cal L}^{(1)}\,+\,\frac{m^2}{2}\,B^{\mu}h_{\mu \nu}B^\nu
\ee
and use the variation of $B_\mu$ and (\ref{varl1}), we get that the total variation vanishes, $\delta {\cal L}^{(2)}\,=\,0$. Let us write down the
explicit form of this action,
\bq
\label{master}
&&{\cal L}^{(2)}= -\frac 14 F_{\mu \nu}F^{\mu \nu} - \frac \rho2 \left(b_\mu F^{\mu \nu}\right)^2 + \frac {m^2}{2}A^\mu h_{\mu \nu} A^\nu  + \nonumber \\
&&- J_\mu B^\mu + \frac{m^2}{2}\,B^{\mu}h_{\mu \nu}B^\nu.
\eq
After carrying out the variation of this action with relation to $B_\mu$, we get the following equation of motion:
\be
J_\mu-m^2h_{\mu \nu} B^\nu=0.
\ee
Plugging this back into (\ref{master}), we obtain a remarkable gauge invariant theory :
\bq
&&{\cal L}_D= \frac 14 F_{\mu \nu}F^{\mu \nu} + \frac {\rho}{2} \left( b_\mu F^{\mu \nu} \right)^2 - \frac {1}{2\alpha} \left( \partial_\mu A^\mu\right)^2
-\frac {1}{2m^2}\left(\partial^\alpha F_{\alpha \mu}\right)\left(\partial^\rho F_{\rho \nu}\right)L^{\mu \nu} + \nonumber \\
&&-\frac{\rho}{m^2}b_\sigma b^\rho \left(\partial^\sigma F_{\rho \nu}\right)\left(\partial^\alpha F_{\alpha \mu}\right)L^{\mu \nu}
+\frac{\rho}{m^2}b_\nu b^\rho \left(\partial^\sigma F_{\rho \sigma}\right)\left(\partial^\alpha F_{\alpha \mu}\right)L^{\mu \nu} + \nonumber \\
&&- \frac{\rho^2}{2m^2}b_\beta b^\alpha b_\sigma b^\rho \left(\partial^\beta F_{\alpha \mu}\right)\left(\partial^\sigma F_{\rho \nu}\right)L^{\mu \nu}
-\frac{\rho^2}{2m^2}b_\mu b_\nu b^\alpha b^\rho \left(\partial^\beta F_{\alpha \beta}\right)\left(\partial^\sigma F_{\rho \sigma}\right)L^{\mu \nu},
\eq
where we have introduced the gauge fixing term, $- \frac {1}{2\alpha} \left( \partial_\mu A^\mu\right)^2$, and
\be
L^{\mu \nu}=\left(h^{-1}\right)^{\mu \nu}= g^{\mu \nu} + \frac{\beta}{1-\beta b^2} b^\mu b^\nu.
\ee
We should also observe that, as it is characteristic of this procedure, higher derivative terms have been generated (some higher dimension operators have been classified in \cite{Pospelov}).

We would like now to study the spectrum of this dual model. Our aim is to check whether the two models are really equivalent or not. As shown in \cite{scarp-epl}, when the only gauge violating term is Procca-like, there will appear in the dual model, besides the excitations present in the original theory, the excitations of a massless model. This massless model is simply the original model without the Procca mass. Yet, this massless excitation appears with the wrong sign. This can spoil the model with ghosts if these nonphysical particles couple to the physical sector. In the present model, we have, besides the Procca mass, a Lorentz-violating mass term. This case is out of the scope of the relation between propagators obtained in \cite{scarp-epl}. So, we intend to verify how the relation proved in \cite{scarp-epl} is modified and if this modification creates new difficulties in the physical interpretation of the dual model.

We will first obtain the propagator of the model described by ${\cal L}_D$. After some partial integrations, we find
\bq
&&{\cal L}_D=\frac 12 A^\mu \left\{ -\frac{1}{m^2}\left( \Box +m^2+ \rho \lambda^2 \right)\left(\Box +\rho \lambda^2\right)\theta_{\mu \nu}
\right. \nonumber \\
&&\left. +\left[\frac{\Box}{\alpha}-\frac{\rho \lambda^2}{m^2}\left(\rho \lambda^2+m^2\right)-\frac{\lambda^2 \Box}{m^2} \frac{1}{1-\beta b^2} \left(\beta+2\rho+\rho^2 b^2\right)
 \right] \omega_{\mu \nu} \right. \nonumber \\
&&\left. +\left[ -\frac{\rho \Box}{m^2}\left(\rho \lambda^2+m^2\right)- \frac{\Box^2}{m^2}\frac{1}{1-\beta b^2}
\left(\beta+2\rho+\rho^2 b^2\right)\right] \Lambda_{\mu \nu} \right. \nonumber \\
&&\left.+ \left[\frac{\rho \lambda}{m^2}\left(\rho \lambda^2+m^2\right)+\frac{\lambda \Box}{m^2}\frac{1}{1-\beta b^2} \left(\beta+2\rho+\rho^2 b^2\right)\right]   \left( \Sigma_{\mu \nu} +\Sigma_{\nu \mu} \right)\right\}A^\nu,
\eq

Using the algebra of Table 1, we can calculate the propagator, which will be given by
\bq
&&\langle A_\mu A_\nu \rangle_D=\frac{i}{A_1 A_2 G}\left\{-m^2 G \theta_{\mu \nu} + \left(\frac{\alpha A_1 A_2 G}{\Box}+\lambda^2m^2F \right) \omega_{\mu \nu} +
\right. \nonumber \\
&& \left.+m^2\Box F \Lambda_{\mu \nu} - \lambda m^2 F \left( \Sigma_{\mu \nu}+\Sigma_{\nu \mu} \right)\right\},
\label{dual}
\eq
with
\bq
&&A_2=\Box + \rho \lambda^2, \\
&&F=\rho A_1+(\rho +\beta) \frac{(1+\rho b^2)}{(1-\beta b^2)} \Box, \\
&&G=\frac{(1+\rho b^2)}{(1-\beta b^2)} \Box E \\
\eq
and where $A_1$ and $E$ are defined in equations (\ref{A1})-(\ref{E}).

We next analyze the residue in the poles to check if the models are classically equivalent. We will consider the situation in which the Lorentz-violating parameter is space-like ($b^\mu=(0,0,0,t)$). It will give us 4 poles:

\begin{eqnarray}
k_{0}^{2} &=& \bold k^2-\rho t^2k_{3}^{2}+m^2\equiv m_1'^2,\\
k_{0}^{2} &=& \bold k^2+\beta t^2k_3^2+\frac{(1+\beta t^2)}{(1-\rho t^2)}m^2\equiv m_2'^2,\\
k_0^2 &=& \bold k^2 \equiv m_3'^2,\\
k_0^2 &=& \bold k^2-\rho t^2k_3^2 \equiv m_4'^2.
\end{eqnarray}
The two first poles are exactly the same of the original model, with $m_1'=m_1$ and $m_2'=m_2$. However, two new modes appear. For the space-like $b_\mu$ defined in (\ref{spacelike}) and again considering the particular situation where $k^\mu=(k^0,0,0,k^3)$, the dual propagator reduces to:

\begin{eqnarray}
\label{dualfinal}
&&\left\langle A_{\mu}A_{\nu}\right\rangle_D =\frac{i}{D_D}\frac{1}{(1-\sigma t^2)}\left\{-m^2 k^2 (1-\sigma t^2)(k_0^2 - m_2'^2)\theta_{\mu\nu}\right.\nonumber\\
&& +\left[-\alpha (1-\sigma t^2)(k_0^2 -m_1'^2)(k_0^2 -m_2'^2)(k_0^2 -m_4'^2)+m^2 t^2k_3^2 F\right]\omega_{\mu\nu}\nonumber\\
&&\left.+m^2k^2t^2F\delta_{\mu}^{3}\delta_{\nu}^{3}+m^2t^2k_3 F(\delta_{\mu}^{3}k_{\nu}+\delta_{\nu}^{3}k_{\mu})\right\},
\end{eqnarray}
with
\be
D_D=(k_0^2 -m_1'^2)(k_0^2 -m_2'^2)(k_0^2 -m_3'^2)(k_0^2 -m_4'^2)
\ee
and
\be
1 - \sigma t^2 = \frac {(1-\rho t^2)^2}{(1+ \beta t^2)}.
\ee
Let us analyze the poles which are common with the original model. For the pole $m_1'^2$, the residue matrix is given by
\be
R'_{1}=
\left(
\begin{array}{cccc}
0& 0 & 0 & 0 \\
0 & 1 & 0 & 0 \\
0 & 0  & 1  & 0 \\
0 & 0 & 0 & 0
\end{array}
\right)
\ee
with eigenvalues

\bq
\lambda _1 &=& 0,  \\
\lambda _2 &=& 1,  \\
\lambda _3 &=& 1,  \\
\lambda _4 &=& 0.
\eq

For the pole $m_2'^2$, we have

\be
R_{2}=\frac{(1+\beta t^2)}{(1-\rho t^2)^2}\frac{m^2}{(m_2'^2-k_3^2)^2}
\left(
\begin{array}{cccc}
k_3^2& 0 & 0 & |m_2'|k_3 \\
0 & 0 & 0 & 0 \\
0 & 0  & 0  & 0 \\
|m_2'|k_3 & 0 & 0 & m_2'^2
\end{array}
\right)
\ee
with eigenvalues

\bq
\lambda _1 &=& 0,  \\
\lambda _2 &=& 0,  \\
\lambda _3 &=& 0,  \\
\lambda _4 &=& \frac{(1+\beta t^2)m^2}{[(1-\rho t^2)\beta t^2k_3^2 + (1+\beta t^2)m^2]}\left[(2+\beta t^2)k_3^2 + \frac{(1+\beta t^2)}{(1-\rho t^2)}m^2\right].
\eq
We see that although the first pole exhibits exactly the same positive eigenvalues for the residue matrix, the second pole gives a different eigenvalue. It is also physically meaningful if we adopt the condition $t^2<<1$ for the cases we studied in the last section, but it appears to receive a contribution from another sector. We shall further discuss this point in the next section, remembering that for the $k^\mu$ we are using, this second pole corresponds to a longitudinal mode of propagation. We will also discuss the origin of the two extra poles in the next section.

\section{General discussion on NDM with Lorentz-violating mass}

As we have seen, the method consists in introducing an auxiliary field, $B_\mu$, such that $\delta B_\mu=
\delta A_\mu= \partial_\mu \eta$, in order to restore gauge invariance, which, in the case of our mass term, will give
\be
{\cal L}_D={\cal L}-J_\mu B^\mu +\frac {m^2}{2}B^\mu h_{\mu \nu} B^\nu.
\ee
In the equation above, $\delta {\cal L}=J_\mu \delta A^\mu$, and $\delta J^\mu=m^2 h^{\mu \nu} \delta A_\nu$,
by virtue of the presence of the mass terms. So, the variation of the Lagrangian ${\cal L}_D$ with respect to $B_\mu$ leads us to
\be
B^\alpha=\frac 1{m^2}L^{\alpha \mu} J_\mu,
\ee
with $L_{\mu \nu}=\left(h^{-1}\right)_{\mu \nu}$, and so
\be
{\cal L}_D={\cal L}-\frac 1{2m^2}JLJ.
\ee
Here we are omitting the Lorentz indices for the sake of simplicity. Let us then consider the two Lagrangians:
\be
{\cal L}=\frac 12 A{\cal O}A
\ee
and
\be
{\cal L}_0=\frac 12 A{\cal O}_0A,
\ee
obtained after suitable partial integrations in their respective actions. In the equation above, ${\cal O}$ and ${\cal O}_0$ are
differential (local) operators corresponding to the theories with or without the mass terms, respectively, which fulfill the relation
\be
{\cal O}={\cal O}_0 + m^2 h.
\label{p1}
\ee

By applying now the NDM to the Lagrangian ${\cal L}$, we have
\be
\delta {\cal L}=\frac 12 \left \{ ({\cal O}A)\delta A + A({\cal O}\delta A)\right \}
= {\cal O}A \delta A.
\label{cor}
\ee
The last step is carried out with the help of partial integration. The differential operator ${\cal O}$  has second-order derivatives or, like in the topological case, a first-order  derivative contracted with the Levi-Civita  tensor density. The equation above allows us to identify the Noether current as
\be
J={\cal O}A.
\ee
Now, we have to add to the Lagrangian the current-current term
\be
-\frac 1{2m^2}JLJ=-\frac 1{2m^2}({\cal O}AL{\cal O}A)= -\frac 1{2m^2}A{\cal O}L{\cal O}A,
\ee
where we again performed partial integrations. Our lagrangian density, then, becomes
\be
{\cal L}_D= \frac 12 A\left ( {\cal O}- \frac 1{m^2}{\cal O}L {\cal O} \right)A.
\ee
The complete dual wave operator can be finally identified as
\be
{\cal O}_D= {\cal O}- \frac 1{m^2}{\cal O}L {\cal O} = {\cal O}\left (
g- \frac 1{m^2}L{\cal O} \right ).
\label{ototal1}
\ee

Using equation (\ref{p1}), we have
\be
g- \frac 1{m^2}L{\cal O} =- \frac 1{m^2}L{\cal O}_0,
\ee
so that
\be
{\cal O}_D=- \frac {1}{m^2}{\cal O}L{\cal O}_0.
\ee

The propagator, as we know, is given by $\langle A_\mu A_\nu \rangle_D = i\left( {\cal O}_D^{-1}\right )_{\mu \nu}$. It is not difficult to invert ${\cal O}_D$, if we know the inverses of ${\cal O}$ and ${\cal O}_0$. For the  ${\cal O}_0$-operator, it is necessary to add a gauge-fixing term, $\frac {\Box}{\alpha} \omega_{\mu \nu}$. Therefore,
\be
{\cal O}_D=- \frac {1}{m^2}{\cal O}L\tilde{\cal O}_0,
\ee
with
\be
\tilde {\cal O}_0= {\cal O}_0 + \frac {\Box}{\alpha} \omega.
\ee
The inverse operator can then be readily written as
\be
{\cal O}_D^{-1}= -m^2 \tilde {\cal O}_0^{-1}h{\cal O}^{-1}.
\ee

Now, we wish to show that this inverse operator is obtained from a simple relation between the inverse operators of the original models. In order to do this, we use that
\be
{\cal O}=\tilde {\cal O}_0 +m^2 h - \frac{\Box}{\alpha}
\omega,
\ee
and then multiply both sides of the equation above at the right hand side by ${\cal O}^{-1}$ and at the left by  $\tilde {\cal O}_0^{-1}$ to obtain
\be
\tilde{\cal O}_0^{-1}= {\cal O}^{-1} + m^2 \tilde {\cal O}_0^{-1}h{\cal O}^{-1} -\frac
{\Box}{\alpha} \tilde {\cal O}_0^{-1} \omega {\cal O}^{-1}
\ee
or
\be
{\cal O}_D^{-1}={\cal O}^{-1}-\tilde {\cal O}_0^{-1}-\frac {\Box}{\alpha} \tilde {\cal O}_0^{-1} \omega {\cal O}^{-1}.
\label{inverso1}
\ee
On the other hand, equation (\ref{ototal1}) can also be written as
\bq
&&{\cal O}_D= {\cal O}- \frac 1{m^2}{\cal O}L {\cal O} = \left (
g- \frac 1{m^2}{\cal O}L \right ){\cal O} \nonumber \\
&&=-\frac{1}{m^2}{\cal O}_0 L {\cal O},
\label{ototal2}
\eq
so that we also have
\be
{\cal O}_D^{-1}={\cal O}^{-1}-\tilde {\cal O}_0^{-1}-\frac {\Box}{\alpha} {\cal O}^{-1} \omega \tilde{\cal O}_0^{-1}.
\ee
The dual total operator can thus be written in the following symmetric form
\be
{\cal O}_D^{-1}={\cal O}^{-1}-\tilde {\cal O}_0^{-1}-\frac 12\frac {\Box}{\alpha}
\left({\cal O}^{-1} \omega \tilde{\cal O}_0^{-1}+\tilde {\cal O}_0^{-1} \omega {\cal O}^{-1}\right).
\label{inverse}
\ee

Now, the operator $\tilde {\cal O}_{0}^{-1}$ can be split into the form
\be
\tilde {\cal O}_0^{-1}=\tilde {\cal O}_{0Tr}^{-1}+ \frac {\alpha}{\Box} \omega,
\ee
due to gauge invariance, where $\tilde {\cal O}_{0Tr}^{-1}$ is transverse ($\omega \tilde {\cal O}_{0Tr}^{-1}=0$). It is easy to show that the most general transversal differential operator expressed in terms of $\theta$, $\omega$, $\Lambda$, $\Sigma$ and $\Sigma^T$ is given by
\be
A_{Tr}=\alpha_1 \theta -\lambda \alpha_2 \omega -\frac{\Box}{\lambda}\alpha_2 \Lambda +\alpha_2\left(\Sigma+\Sigma^T\right),
\ee
where $\alpha_1$ and $\alpha_2$ are differential operators. From this expression, we can conclude that
\be
{\cal O}^{-1}={\cal O}_{Tr}^{-1 }+ c_1 \omega + c_2 \Lambda.
\label{splitting}
\ee
With the above splitting, using the algebra in Table 1, we have
\be
-\frac 12\frac {\Box}{\alpha}
\left({\cal O}^{-1} \omega \tilde{\cal O}_0^{-1}+\tilde {\cal O}_0^{-1} \omega {\cal O}^{-1}\right)=
-c_1 \omega -\frac 12 \frac{\lambda}{\Box} c_2\left(\Sigma + \Sigma^T\right),
\ee
which, from equation (\ref{inverse}), will give us the following relation between the propagators
\be
\langle A_\mu A_\nu \rangle_D= \langle A_\mu A_\nu \rangle-\langle A_\mu A_\nu \rangle _0
- i c_1 \omega_{\mu \nu} -\frac i2 \frac{\lambda}{\Box} c_2\left(\Sigma_{\mu \nu} + \Sigma_{\nu \mu}\right),
\label{propag}
\ee
with $c_1$ and $c_2$ being differential operators.

We are now in position to analyze the results of the last section. First, we see the reason why we found the poles of the massless model in the dual theory. This occurs because we have the difference between the massive and non massive propagators in equation (\ref{propag}). This also reveals that the poles of the massless model are to be associated with non physical modes, since the corresponding propagator appears with the wrong sign. The relation above is very similar to the one obtained in ref. \cite{scarp-epl}, the manifest difference being the presence of the last term in (\ref{propag}). Finally, the modification of the residue of the second pole of the dual model must come from the two last terms. Let us verify this. From (\ref{splitting}), we have
\be
\omega {\cal O}^{-1}=c_1 \omega +\frac{\lambda}{\Box} c_2 \Sigma^T.
\ee
It is an easy task to show that for our original model, the operators $c_1$ and $c_2$ are given by
\be
c_1=\frac{1}{m^2} + \beta \frac{\lambda^2\left(1+\rho b^2\right)}{m^2 E}
\ee
and
\be
c_2=\frac{\beta}{E},
\ee
which, using the relation in equation (\ref{propag}), will give us
\bq
&&\langle A_\mu A_\nu \rangle_D= \langle A_\mu A_\nu \rangle-\langle A_\mu A_\nu \rangle _0 +\nonumber \\
&&-i\left(\frac{1}{m^2} + \beta \frac{\lambda^2\left(1+\rho b^2\right)}{m^2 E}\right)\omega_{\mu \nu}
-i\beta \frac{\lambda}{2 \Box E} \left(\Sigma_{\mu \nu} + \Sigma_{\nu \mu} \right),
\eq
where there is a difference of sign in the gauge-fixing term which will be explained below.
It is clear in the equation above that the last two terms affect the residue of the second pole, containing in the denominator a factor $E$, although this pole is still physically meaningful. Moreover it is also evident that in the limit $\beta \to 0$ we recover the relation of \cite{scarp-epl}.

Nevertheless, it appears we have another problem. If we take explicitly the two propagators $\langle A_\mu A_\nu \rangle$ and $\langle A_\mu A_\nu \rangle_0$, with the propagator for the massless theory given by
\bq
&&\langle A_\mu A_\nu \rangle_0= \frac{i}{\Box A_2}\left\{ \Box \theta_{\mu \nu} +\left( \alpha A_2 -\frac{\rho \lambda^2}{(1+b^2 \rho)}\right)\omega_{\mu \nu}
-\frac{\rho \Box}{(1+b^2 \rho)} \Lambda_{\mu \nu} \right. \nonumber \\
&&\left.+\frac{\rho \lambda}{(1+b^2 \rho)}\left(\Sigma_{\mu \nu}+\Sigma_{\nu \mu}\right)\right\},
\eq
we obtain
\be
c'_1=\frac{1}{m^2} + \beta \frac{\lambda^2\left(1+\rho b^2\right)}{m^2 E}-\frac{\beta \lambda^2}{\Box E} - 2\frac{\alpha}{\Box}
\ee
and
\be
c'_2=2\frac{\beta}{E}.
\ee
This apparent contradiction is actually due to two different procedures. In section III, we obtained the dual lagrangian density, added a gauge fixing term and then inverted the wave operator. To arrive in relation (\ref{propag}), on the other hand, we added the gauge fixing term to the the wave operator ${\cal O}_0$, which was a factor of ${\cal O}_D$. If we calculate ${\cal O}_D$ explicitly by using the known operators $\tilde{\cal O}_0$ and ${\cal O}$, we will see that the net effect of this different approach is the use of a new gauge fixing directly in ${\cal O}_D$, given by
\be
{\cal O}_{GF}=-\frac{\Box}{\alpha}\left\{\left[1- \beta \frac{\lambda^2}{m^2} \frac{(1+\rho t^2)}{(1-\beta t^2)}\right]\omega_{\mu \nu}
+\frac{\beta \lambda}{2m^2}\frac{(1+\rho t^2)}{(1-\beta t^2)}\left(\Sigma_{\mu \nu}+\Sigma_{\nu \mu}\right)\right\}.
\ee
As a result, the dual propagator acquired new contributions in the $\omega$ and $\Sigma + \Sigma^T$ sectors. The calculation following this second procedure furnishes us
\bq
&&\langle A_\mu A_\nu \rangle'_D=\frac{i}{A_1 A_2 G}\left\{-m^2 G \theta_{\mu \nu} + \left[-\frac{\alpha A_1 A_2 G}{\Box}+\lambda^2m^2F
-\beta \lambda^2 A_1 A_2 \frac{(1+\rho b^2)}{(1-\beta b^2)}\right] \omega_{\mu \nu} +
\right. \nonumber \\
&& \left.+m^2\Box F \Lambda_{\mu \nu}
+\left[-\lambda m^2 F + \frac{\beta \lambda A_1 A_2}{2} \frac{(1+\rho b^2)}{(1-\beta b^2)}\right]\left( \Sigma_{\mu \nu}+\Sigma_{\nu \mu} \right)\right\}.
\eq

A comment is in order. First, the difference in the propagators for the dual model obtained in the previous section and in the present one is due to the use of different gauge fixing terms and, so, this has no physical consequence. For the particular case where the electromagnetic wave propagates parallel to the background vector, these differences occur in the longitudinal mode, without affecting the transversal mode. However, the mode corresponding to the pole $m_2'^2$ is longitudinal only in this particular situation.

Concerning the presence of ghost modes, we see in the context we have investigated that they spoil the present dual model if they couple to some physical sector. However, it is interesting to verify if in a quantum calculation these non physical contributions could be decoupled from the physical ones. It is known that the dualization procedure, when interactions are considered, originates nonminimal couplings. These new couplings should be taken into account in a future investigation.

\section{Concluding comments}

We analyzed the gauge sector of a CPT-even model with a Lorentz-breaking mass term. The model was shown to present physical massive poles which, depending on the choice of the coefficients $\rho$ and $\beta$, have its degrees of freedom changed. For this class of models, interesting physical aspects can be accommodated for particular values of $\rho$ and $\beta$, like, for example, the presence of propagating superluminal modes.

We also dedicated attention to the search of a physical equivalent gauge invariant model. The theory was gauge embedded with the use of the Noether Dualization Method (NDM). It was verified that the dual model presents, besides the original massive physical modes, two more poles coming from the massless version of the original theory. These two new poles were shown to be non physical, characterizing ghosts. These ghosts spoil the dual model if they couple to the physical sector. It is interesting to verify for the present photon dual lagrangian density, when considered in a complete model, if in a quantum calculation of some process these non physical contributions could be decoupled from the physical ones. This would be in accordance with the belief that the non-invariant model can be considered as the gauge fixed version of a gauge theory. Besides, it is known that the dualization procedure, when interactions are considered, originates nonminimal couplings. These new couplings should be taken into account in a future investigation.

\vspace{5mm}
\noindent
{\bf Acknowledgments}

The authors acknowledge fruitful discussions with prof. J. A. Helay\"el-Neto . M. S. and A. P. B. S. acknowledge research grants from CNPq. H. G. Fargnoli thanks CAPES/FAPEMIG for financial support.


\begin{thebibliography}{99}

\bibitem{kostelecky1}  D. Colladay and V. A. Kostelecky, \textit{{Phys. Rev.
\textbf{D}}} \textbf{55}, (1997) 6760.

\bibitem{kostelecky2}  D. Colladay and V. A. Kostelecky, \textit{{Phys. Rev.
\textbf{D}}} \textbf{58}, (1998) 116002.

\bibitem{coleman1}  S. Coleman and S. L. Glashow, \textit{{Phys. Lett. \textbf{B}}}
\textbf{405}, (1997) 249.

\bibitem{coleman2} S. Coleman and S. L. Glashow, \textit{{Phys. Rev.
\textbf{D}}} \textbf{59}, (1999) 116008.

\bibitem{data-exp} V. A. Kostelecky and Neil Russell, \textit{{Rev. Mod. Phys.}} \textbf{83}, (2011) 11.

\bibitem{Carroll} S. M. Carroll, G. B. Field and R. Jackiw, \textit{Phys. Rev. \textbf{D}}{\bf 41}, (1990) 1231.

\bibitem{Goldhaber} M. Goldhaber and V. Trimble, {\it J. Astrophys. Astron.}{\bf 17}, (1996) 17.

\bibitem{CS1} R. Jackiw and V. A. Kostelecky, \textit{{Phys. Rev. Lett.}} \textbf{82}, (1999) 3572.

\bibitem{CS2} M. Perez-Victoria, {\it Phys. Rev. Lett.} {\bf 83}, 2518 (1999).

\bibitem{CS3} J.M. Chung and P. Oh, \textit{Phys. Rev. \textbf{D}}{\bf 60}, 067702 (1999).

\bibitem{CS4} A. P. Ba\^eta Scarpelli, M. Sampaio, M. C. Nemes, and B. Hiller,
\textit{{Phys. Rev. \textbf{D}}} \textbf{64}, (2001) 046013.

\bibitem{CS5} M. Perez-Victoria, \textit{{J. High. Energy Phys.}} \textbf{0104}, 032 (2001).

\bibitem{CS6} B. Altschul, \textit{Phys. Rev. \textbf{D}}{\bf 69}, 125009 (2004).

\bibitem{CS7} B. Altschul, \textit{{Phys. Rev. \textbf{D}}} \textbf{70}, (2004) 101701.

\bibitem{CS8}  A.P. Ba\^eta Scarpelli, M. Sampaio, M.C. Nemes, B. Hiller, \textit{{Eur. Phys. J. \textbf{C}}} \textbf{56}, (2008) 571.

\bibitem{Carroll2} S. Carroll and H. Tam, \textit{{Phys. Rev. \textbf{D}}} \textbf{78}, (2008) 044047.

\bibitem{NM-coupling} M. Gomes, J. R. Nascimento, A. Yu. Petrov, A. J. da Silva, \textit{{Phys. Rev. \textbf{D}}} \textbf{81}, (2010) 045018;
G. Gazzola, H.G. Fargnoli, A.P. Ba\^eta Scarpelli, Marcos Sampaio, M.C. Nemes, \textit{{J. Phys. \textbf{G}}} \textbf{39} (2012) 035002;
A.P. Ba\^eta Scarpelli, \textit{{J. Phys. \textbf{G}}} \textbf{39} (2012) 125001;
A. P. Ba\^eta Scarpelli. T. Mariz, J. R. Nascimento, A. Yu. Petrov, \textit{{Eur. Phys. J. \textbf{C}}} \textbf{73}, (2013) 2526.

\bibitem{belich-mmferreira} H. Belich, T. Costa-Soares, M. M. Ferreira Jr., J.A. Helayel-Neto, \textit{{Eur. Phys. J. \textbf{C}}} \textbf{41}, (2005) 421;
H. Belich, T. Costa-Soares, M. M. Ferreira Jr., J.A. Helayel-Neto, \textit{{Eur. Phys. J. \textbf{C}}} \textbf{42}, (2005) 127;
H. Belich, T. Costa-Soares, M. M. Ferreira Jr., J.A. Helayel-Neto, F. M. O Moucherek, \textit{{Phys. Rev. \textbf{D}}} \textbf{74}, (2006) 065009;
H. Belich, L.P. Colatto, T. Costa-Soares, J.A. Helayel-Neto, M.T.D. Orlando, \textit{{Eur. Phys. J. \textbf{C}}} \textbf{62}, (2009) 425.

\bibitem{shan-quan1} Shan-quan Lan, Feng Wu, \textit{Phys.Rev. \textbf{D}} {\bf 87} (2013) 12, 125022.

\bibitem{kost-5d} Alan Kostelecky, Matthew Mewes, \textit{Phys.Rev. \textbf{D}} {\bf 88} (2013) 9, 096006.

\bibitem{shan-quan2} Shan-quan Lan, Feng Wu, \textit{{Eur. Phys. J. \textbf{C}}} \textbf{74}, (2014) 2875.

\bibitem{lv-mass1} G. Gabadadze, L. Grisa, \textit{Phys. Lett. \textbf{B}} {\bf 617}, (2005) 124.

\bibitem{lv-mass2} G. Dvali, M. Papucci, M. D. Schwartz, \textit{Phys. Rev. Lett.} {\bf 94}, (2005) 191602.

\bibitem{lv-mass3} B. Altschul, \textit{Phys. Rev. \textbf{D}} {\bf 73}, (2006) 036005.

\bibitem{lv-mass4} M. Cambiaso, R. Lehnert, R. Potting, \textit{{Phys. Rev. \textbf{D}}} {\bf 85}, (2012) 085023.

\bibitem{lv-mass5} Brett Altschul, \textit{Phys.Rev.\textbf{D}} {\bf 86}, (2012) 045008.

\bibitem{LCTBrito} L. C. T. Brito, H. G. Fargnoli, A.P. Baêta Scarpelli, \textit{Phys.Rev. \textbf{D}} {\bf 87} (2013) 125023.

\bibitem{Deser-Jackiw} S. Deser, R. Jackiw, \textit{Phys. Lett.\textbf{B}} {\bf 139}, (1984) 371.

\bibitem{Kenneth} Kenneth A. Intriligator, N. Seiberg, \textit{Nucl.Phys.Proc.Suppl}. {\bf 45BC} (1996) 1-28.

\bibitem{review} S. E. Hjelmeland and U. Lindstrom, hep-th/9705122.

\bibitem{Ilha1} M. A. Anacleto, A. Ilha, J. R. S. Nascimento, R. F. Ribeiro and C. Wotzasek, \textit{Phys.
Lett. \textbf{B}} {\bf 504} (2001) 268.

\bibitem{Ilha2} A. Ilha and C. Wotzasek, \textit{Nucl. Phys. \textbf{B}} {\bf 604} (2001) 426.

\bibitem{Nieuw} D. Z. Freedman and P. van Nieuwenhuizen, \textit{Phys. Rev. \textbf{D}} {\bf 13} (1976) 3214.

\bibitem{Ferrara1} S. Ferrara, D. Z. Freedman and P. van Nieuwenhuizen, \textit{Phys. Rev. \textbf{D}} {\bf 15} (1977) 1013.

\bibitem{Ferrara2} S. Ferrara and J. Scherk, \textit{Phys. Rev. Lett.} {\bf 37} (1976) 1035.

\bibitem{scarp-epl} A.P. Baeta Scarpelli, M. Botta Cantcheff and J.A. Helayel-Neto, \textit{Europhys. Lett.}\textbf{65}, (2004) 760-765.

\bibitem{Bota} M. Botta Cantcheff, C. F.  L. Godinho, A.P. Baeta Scarpelli, J.A. Helayel-Neto, \textit{Phys. Rev. \textbf{D}} {\bf 68} (2003) 065025.

\bibitem{Petrov} M.S. Guimaraes, J.R. Nascimento, A.Yu. Petrov, C. Wotzasek, \textit{Europhys. Lett.} {\bf 95} (2011) 51002.

\bibitem{Dalmazi1} D. Dalmazi, \textit{JHEP} {\bf 0601} (2006) 132.

\bibitem{Dalmazi2} D. Dalmazi, \textit{JHEP} {\bf 0608} (2006) 040.

\bibitem{Dalmazi3} D. Dalmazi, Elias L. Mendon\c{c}a, \textit{J. Phys. \textbf{A}} {\bf 39} (2006) 11091-11099.

\bibitem{Dalmazi4} D. Dalmazi, Elias L. Mendon\c{c}a, \textit{JHEP} {\bf 0909} (2009) 011.

\bibitem{Belich-spectrum} H. Belich, F. J. L. Leal, H. L. C. Louzada, M. T. D. Orlando, \textit{Phys. Rev. \textbf{D}} {\bf 86} (2012) 125037.

\bibitem{Pospelov} Pavel A. Bolokhov, Maxim Pospelov, \textit{Phys. Rev. \textbf{D}} {\bf 77} (2008) 025022.



\end{thebibliography}
\end{document}